\documentclass[aps,prl,onecolumn,groupedaddress]{revtex4}
\usepackage[draft]{hyperref}
\usepackage{amsmath,amssymb}
\usepackage{graphicx}
\usepackage{color}

\begin{document}

\title{Soliton-induced relativistic-scattering and amplification.}

\author{E.~Rubino,$^{1}$  A.~Lotti,$^{1,2}$ F. Belgiorno,$^{3}$ S.~L.~Cacciatori,$^{1}$ A.~Couairon,$^{2}$, U.~Leonhardt,$^{4}$, and D.~Faccio$^{5,*}$}

\address{
$^{1}$ Dipartimento di Scienza e Alta Tecnologia, Universit\`a dell'Insubria, Via Valleggio 11, IT-22100 Como, Italy\\
$^{2}$ Centre de Physique Theorique CNRS, Ecole Polytechnique, F-91128 Palaiseau, France\\
$^{3}$ Dipartimento di Matematica, Politecnico di Milano, Piazza Leonardo 32,20133 Milano, Italy\\
$^{4}$ School of Physics and Astronomy, University of St Andrews, North Haugh, St Andrews KY16 9SS, UK\\
$^{5}$ School of Engineering and Physical Sciences, SUPA, Heriot-Watt University, Edinburgh EH14 4AS, UK
}
* Correspondence and requests for materials should be addressed to D.F. ({d.faccio@hw.ac.uk}).

\maketitle

{\bf Solitons are of fundamental importance in photonics due to applications in optical data transmission and also as a tool for investigating novel phenomena ranging from light generation at new frequencies and wave-trapping to rogue waves. Solitons are also relativistic scatterers: they generate refractive-index perturbations moving at the speed of light. Here we found that such perturbations scatter light in an unusual way: they amplify light by the mixing of positive and negative frequencies, as we describe using a first Born approximation and numerical simulations. The simplest scenario in which these effects may be observed is within the initial stages of optical soliton propagation: a steep shock front develops that may efficiently scatter a second, weaker probe pulse into relatively intense positive and negative frequency modes with amplification at the expense of the soliton. Our results show a novel all-optical amplification scheme that relies on relativistic scattering.}

If compared to the well-developed field of traditional light scattering in which the medium is at rest, little attention has been devoted to the physics of scattering from a moving medium, in particular from a relativistically moving medium. 
Here we consider the remarkable ability of solitons to generate a co-propagating refractive index inhomogeneity that propagates at relativistic speeds. The basics of scattering from a time-changing boundary were discussed in detail by {Mendon\c{c}a} and co-workers (see e.g. \cite{mendonca} and references therein). Examples of such ``time refraction'' have been predicted and observed from a moving plasma front \cite{mendonca,rosanov,mori} and in waveguide structures \cite{gaburro,lipson,krauss}. Recently, the nonlinear Kerr effect, i.e. the local increase of the medium refractive index induced by an intense laser pulse \cite{boyd}, was proposed to induce a  moving refractive index inhomogeneity within a dispersive medium such as an optical fibre \cite{ulf}. The laser pulse induced relativistic inhomogeneity (RI) was then described in terms of a  flowing medium in which the analogue of an event horizon may form \cite{ulf} and applications such a optical transistors have been proposed \cite{steinmeyer}.\\
Intense laser pulses are also known to scatter from the self-induced travelling RI: this self-scattering process leads to the resonant transfer of energy from the laser pulse to a significantly blue-shifted peak, often referred to as resonant radiation (RR) or ``optical Cherenkov'' radiation \cite{cerenkov,wai,cristiani,skryabin,dudley}. A recent discovery highlighted an additional scattered mode, further blue-shifted with respect to the RR, identified as a mode excited on the negative frequency branch of the medium dispersion relation and therefore named ``negative resonant radiation'' (NRR) \cite{NRR}.\\
\begin{figure}[b]
\centering
\includegraphics[width=8cm]{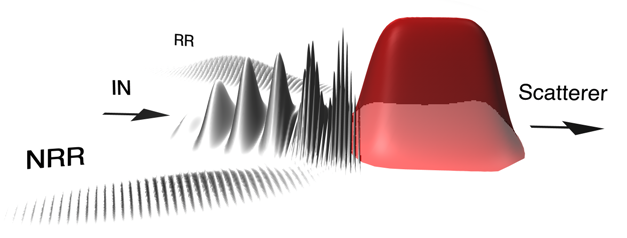}
\caption{Schematic representation of scattering from a relativistic inhomogeneity:  an incoming laser pulse, IN, interacts with a co-propagating RI, or scatterer, that transfers energy into two output modes, RR and NRR. }
\label{fig:scheme}
\end{figure}
By applying a first Born approximation analysis, supported by numerical simulations, we show that the scattering of light by a soliton induced RI reveals a novel all-optical amplification mechanism. A schematic representation of the specific process we are considering is shown in Fig.~\ref{fig:scheme}: an incoming laser pulse, IN, interacts with a co-propagating inhomogeneity, or scatterer. 
The momentum conservation law that governs the scattering process predicts that light may resonantly scatter into two output modes, RR and NRR (see Additional Material).
In the laboratory reference frame, both of these modes will have positive frequencies while in the reference frame comoving with the scatterer, RR is positive and NRR is negative valued. 
The key point in the following is precisely the presence of these two modes: mixing between the RR and NRR modes leads to a novel amplification process.\\ 
\begin{figure*}[t]
\centering
\includegraphics[scale=1]{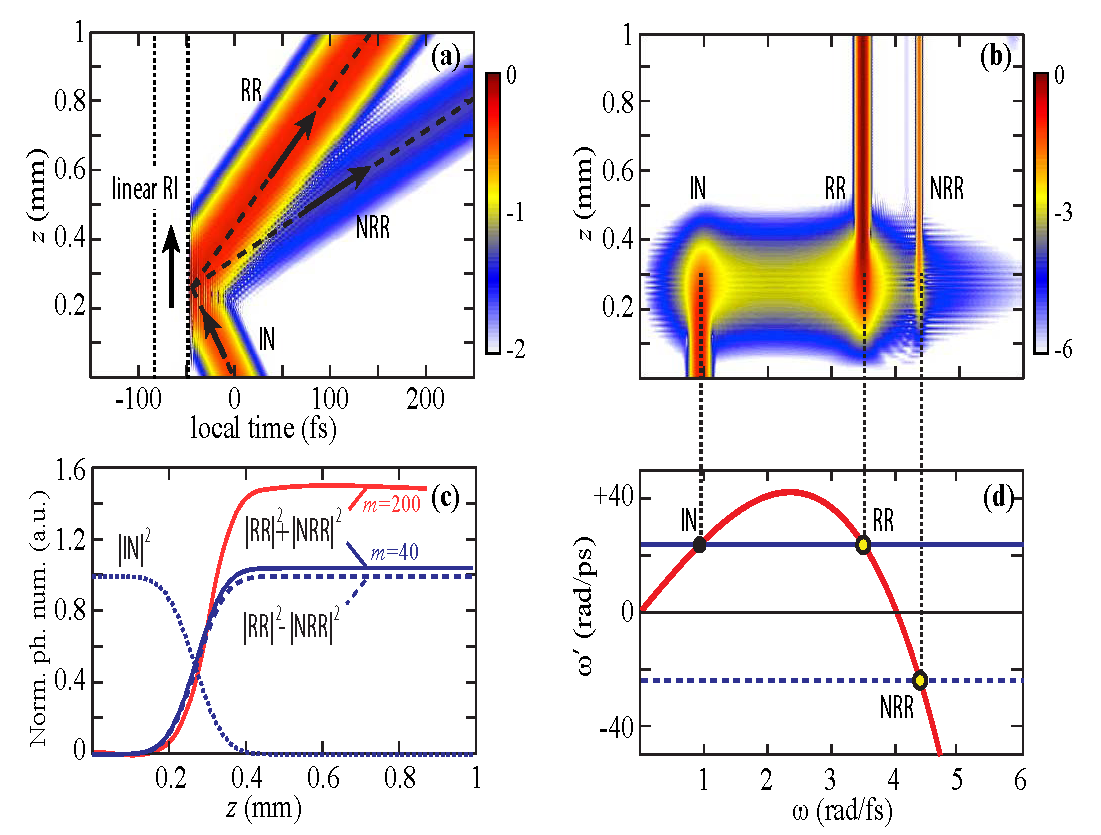}
\caption{Scattering from a ``toy-model'' RI in diamond: IN pulse at $2$ $\mu$m; super-Gaussian moving RI with $m=40$ (3 fs rise-time), $\delta n_{0}=0.08$. Envelope (a) and spectral (b) evolution along propagation. (c) Photon number evolution along propagation: the total sum (photon number amplification) increases with increasing RI steepness (increasing parameter $m$). (d) Comoving dispersion curve with indicated the allowed optical modes.}
\label{fig:esempio}
\end{figure*}
\section{Results}
\noindent{\bf First Born approximation. }
In scattering theory, the first Born approximation allows us to estimate the elements of the scattering matrix $\mathbf{S}$, which connects asymptotic input and output states. Kolesik \emph{et al.} have shown that the positive RR mode is captured in great detail by the Born scattering approach \cite{kolesik-2010}. Here we extend this theory in order to also include the possibility of scattering to the NRR mode and find that the amplitude of each scattering channel is proportional to the amplitude of the Fourier component of the RI, $\hat{R}(\omega)$ (see Additional Material):
\begin{eqnarray}
\label{eq:born}
S(z,\omega_{\rm{RR}}) &\approx& \frac{v}{2} \hat{R}(\omega_{\rm{RR}}-\omega_{\rm{IN}}) 
\end{eqnarray}
and, accounting also for the negative frequency branch, we similarly obtain:
\begin{eqnarray}
S(z,\omega_{\rm{NRR}}) &\approx& \frac{v}{2} \hat{R}(\omega_{\rm{NRR}}+\omega_{\rm{IN}}),
\label{eq:Nborn}
\end{eqnarray}
where $v$ is the velocity of the travelling RI.
These relations (\ref{eq:born}) and (\ref{eq:Nborn}) state that energy may be transferred to \emph{two} modes, RR and NRR, and in order to do so the spectrum of the scattering potential must have non-zero Fourier components 
at frequencies equal to the relative distances between $\omega_{\rm{IN}}$ and the resonant frequencies.
In other words, the RI must present a sufficiently steep gradient so as to effectively excite the desired output modes.
We now observe  
that we may derive a photon number balance equation by generalizing the Manley-Rowe relation, adopted e.g. in nonlinear optics \cite{boyd}, to the case of a {\emph{moving}} scatterer \cite{ostrovskii,kravtsov-ostrovskii-stepanov}. We find  that:
\begin{equation}
|\rm{RR}|^{2}-|\rm{NRR}|^{2}=1,
\label{eq:gen-mr}
\end{equation}
where $|{\rm{RR}}|^{2}$ and $|{\rm{NRR}}|^{2}$ are the photon numbers of the RR and NRR modes normalized to the input photon number, $|{\rm{IN}}|^{2}$. The negative sign in front of the $|{\rm{NRR}}|^{2}$ photon number is a direct consequence of the fact that the ${\rm{NRR}}$-mode has negative frequency in the comoving reference frame (see Additional Material).
So the \emph{difference} between the normalized number of photons has to be equal to the photon number in the input mode. As a consequence, the total output photon number, $|\rm{RR}|^{2}+|\rm{NRR}|^{2}>1$, i.e. we have amplification. 
The scattering process mediated by the  travelling RI  will amplify photons as a result of the coupling between the positive and negative frequency modes.\\

\noindent{\bf Numerical simulations. }
We verified our predictions by  numerically simulating the one dimensional propagation of a scalar field  propagating along the $z$ direction \cite{uppe} in a transparent and isotropic dielectric medium (diamond and fused silica)  that is arranged to scatter from a co-propagating inhomogeneity (see Methods). We consider three separate cases: (i) the underlying physics are first exemplified by taking a ``toy-model'' situation in which the RI is simply a variation in the linear refractive index; (ii) the RI is generated by an actual soliton-like laser pulse through the nonlinear Kerr effect and light from the soliton itself is self-scattered; (iii) the RI is generated by a soliton-like pulse and a second, very weak probe pulse is scattered by the soliton.

{(i) - {\emph{``Toy-model'' RI}.} 
{ The RI is simulated as a linear propagating refractive index inhomogeneity that moves with a speed $v=1.23\times 10^8$ m/s that is just slightly slower than the group velocity of an input probe pulse in the medium, $v_{g}=1.25\times 10^8$ m/s. } 
{In this case, as a dielectric medium we choose diamond, which does not exhibit any resonances over a very
broad bandwidth (from less than one Terahertz through to ultraviolet wavelengths).}
The probe pulse is very weak so that no nonlinear effects are excited and the physics are dominated solely by linear scattering from the RI. Figures~\ref{fig:esempio}(a)-(b) show the evolution along propagation of the envelope (plotted in the frame moving at the RI velocity) and of the spectrum, respectively.
We launch an input Gaussian probe pulse (carrier wavelength $2$ $\mu$m, pulse width $20$ fs), together with a super-Gaussian shaped RI (see Methods), with amplitude {$\delta n_{0}=0.08$} and super-Gaussian order $m=40$ corresponding to a rising time $\sim3$ fs; $\sigma=50$ fs. This ensures a wide RI Fourier spectrum $\widehat{\delta n}(\omega)$, so as to meet both conditions (\ref{eq:born}) and (\ref{eq:Nborn}). 
After $\sim0.2$ mm the probe pulse reaches the moving RI [whose position is highlighted in Fig.~\ref{fig:esempio}(a) by the vertical black dotted lines] and slows down due to the RI increase of refractive index. Scattering occurs towards \emph{two} blue-shifted resonant modes, RR and NRR, which due to dispersion both travel slower than the RI. 
The medium dispersion relation is shown in Fig. \ref{fig:esempio}(d) in  $(\omega^{\prime},\omega)$ coordinates where the comoving frequency $\omega^{\prime}=\gamma(\omega-vk)$; $k=\omega n(\omega)/c$ is the medium dispersion relation and $\gamma=1/\sqrt{1-(v/c)^{2}}$, being $c$ the speed of light in vacuum. In the comoving frame, the absolute value of scattered mode frequencies are determined by the input mode comoving frequency (horizontal solid and dashed lines, see Additional Material): the corresponding laboratory reference frame frequencies can be seen to be in perfect agreement with the numerically observed values.\\
In order to verify the photon number amplification predicted by the Manley-Rowe relation (\ref{eq:gen-mr}), we evaluated the photon number evolution, as shown in Fig.~\ref{fig:esempio}(c) for two different RI gradients, obtained by varying the order of the super-Gaussian function that describes the RI, $m=40$ and $m=200$ (these correspond to rise times from the background refractive index to the maximum of the RI of 3 fs and 0.7 fs, respectively).
Dotted, solid and dashed lines correspond to the normalized photon numbers in the IN mode, total output photons $|\rm{RR}|^{2}+|NRR|^{2}$  and output photon difference $|\rm{RR}|^{2}-|NRR|^{2}$, respectively. As can be seen, the difference in photon numbers is conserved while the sum of photon numbers is larger than $1$ (amplification) and increases with increasing RI steepness. This simplified ``toy-model'' therefore gives a direct confirmation of the predictions based on the first Born approximation model.

\begin{figure}[t]
\centering
\includegraphics[width=8cm]{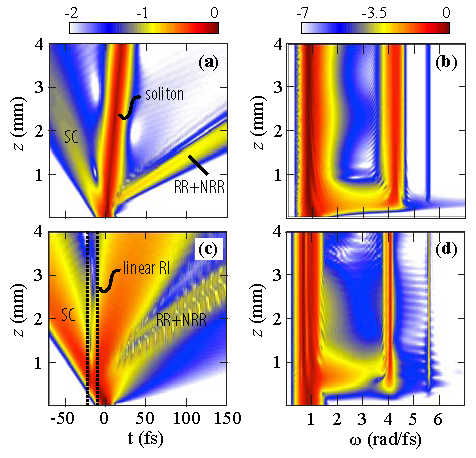}
\caption{RI scattering dynamics in fused silica. Comparison between a soliton-induced (via the \emph{nonlinear} Kerr effect) RI (a)-(b) and the ``toy-model'' RI with the same shape as the soliton-induced inhomogeneity, in a purely linear medium (c)-(d). Envelope (left hand side), and spectral (right hand side) evolution in propagation. 
}
\label{fig:nonlin}
\end{figure}
{(ii) - {\emph{Soliton-induced RI}.} 
The RI is physically generated by an intense  soliton-like laser pulse  through the nonlinear Kerr effect, i.e. $\delta n=n_{2}I$, where $I$ is the peak intensity and $n_{2}$ the nonlinear Kerr coefficient (see Methods).
The  soliton may interact with the self-induced refractive index variation and scatter into new output modes. 
As a Kerr medium we choose common fused silica glass, whose dispersion relation is well known \cite{agrawal1}. We launch a soliton-like mode in the anomalous dispersion region, with central wavelength $2$ $\mu$m, pulse width $7$ fs and input intensity $10$ TW/cm$^{2}$. Figures \ref{fig:nonlin}(a)-(b) show the envelope and spectral evolution, respectively, along propagation over a distance $z=0.4$ cm. 
The input pulse initially self-steepens, forms a shock front and, at $z\sim0.3$ mm, when conditions (\ref{eq:born})-(\ref{eq:Nborn}) are met, energy starts to resonantly transfer from the input mode towards two blue-shifted modes, RR and NRR, both travelling slower with respect to the input soliton. The soliton frequency is red-shifted due to a recoil effect \cite{cerenkov} and subsequently emerges with a slightly lower propagation velocity. 
In Figure \ref{fig:nonlin}(c)-(d) we compare the nonlinear propagation  with a numerical simulation in which the nonlinear source term has been replaced with a RI in the \emph{linear}  polarization with the same amplitude, steepness and propagation velocity of the soliton induced perturbation of Figs.~\ref{fig:nonlin}(a)-(b). 
We observe a nearly perfect agreement with the nonlinear case confirming that soliton shedding of RR and NRR modes is nothing more than a specific realization of the more general relativistic scattering process. 
The ``toy-model'' RI scattering  is indeed able to capture all the essential features of the resonant energy transfer, and coupling between positive and negative frequency modes.

{(iii) - {\emph{Probe pulse scattering from a soliton-induced RI}.} 
Finally, we consider the case in which the soliton is accompanied by a second, delayed probe pulse. This second pulse is much weaker and therefore does not form a soliton or excite any nonlinear Kerr effects.
\begin{figure}[t]
\centering
\includegraphics[width=8cm]{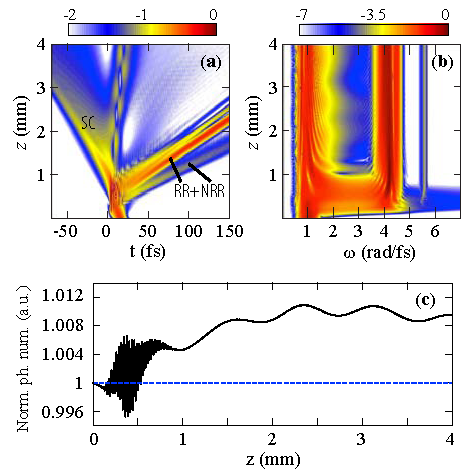}
\caption{Weak probe scattering in fused silica, from the soliton-induced RI of Fig. \ref{fig:nonlin}(a)-(b). Probe envelope (a) and spectral evolution in propagation (b). Normalised photon number evolution (c) for the probe (black solid curve) and for the soliton (blue dashed curve). A complete movie animation is included.
}
\label{fig:probe}
\end{figure}
Figure \ref{fig:probe} shows the results for a $1.9$ $\mu$m central wavelength probe pulse, pulse width $8$ fs and input intensity $5\times10^{9}$ W/cm$^{2}$ (complete movie showing the full dynamics is also included). The probe pulse is scattered by the soliton induced RI shown in Fig. \ref{fig:nonlin}(a)-(b).
The initial probe pulse delay, $t_{0}=15$ fs, is adjusted such that it encounters the soliton when it forms the steepest shock-front and is then scattered relatively efficiently, simultaneously spectrally recoiling to a slightly red-shifted wavelength and emitting two blue-shifted RR and NRR modes. 
In Fig. \ref{fig:probe}(c) we show the overall photon number evolution for the probe pulse (black solid line) and for the soliton pulse (blue dashed curve). In order to highlight the cross-scattering dynamics, the two curves have been normalized at each propagation distance with respect to two independent simulations for the probe and soliton pulses alone.
We thus note that the overall probe pulse photon number increases by nearly $1\%$, followed by some weaker oscillations that are originating from a remnant of the probe pulse that is trapped and thus continuously interacts with the soliton [see Fig.~\ref{fig:probe}(a), trapped light is visible for $z>1.5$ mm, around $t\sim0$ fs]. This clear photon number increase thus indicates that soliton-induced scattering in this regime, very differently from standard cross-phase modulation, may lead to true amplification of the weaker pulse.

\section{Discussion}
\noindent Summarising, we have shown that a RI amplifies and scatters light to higher frequencies. Likewise, if the probe pulse were to be reduced to the level of quantum fluctuations, we may expect to see the RI excite the vacuum states. 
{Bearing in mind that the motion of photons in the vicinity of a RI may also be described in terms of an effective curved space-time metric \cite{visser,faccio1}, we expect that measurements of vacuum fluctuations excited by a RI would give an experimental direct window into the physics of the quantum vacuum in curved space-times \cite{birrell,belgiorno,rubino}.}

These results therefore open the perspective for novel all-optical control schemes that may be implemented in a wide variety of geometries and applications together with a novel numerical and experimental approach for the investigation of fundamental phenomena.

\section{Methods}

\noindent {\bf Numerical model.}
The code adopted for the numerical results presented in the main text is a 1D Unidirectional Pulse Propagation Equation (UPPE) solver \cite{uppe}, where the spectral components $E_{\omega}(z)$ of the real field $E(z,t)$, are obtained by Fourier transform 
and obey to the following model equation, in the pulse or inhomogeneity comoving frame:
\begin{equation}
\partial_{z}{E_{\omega}}-i[k_{z}(\omega)-\omega/v]{E_{\omega}} = i\frac{\omega^{2}}{2 \epsilon_0 c^{2} k_{z}(\omega)}\, P_\omega,
\label{eq:model}
\end{equation}
being $\epsilon_{0}$ the vacuum permittivity, $c$ the speed of light in vacuum, and $P_\omega$ the spectral component at frequency $\omega$ of the real-valued polarization as obtained by Fourier transform.\\
Dispersion of the medium is described by $k_z(\omega)$ and the shift $-\omega/v$ indicates that Eq.~(\ref{eq:model}) is solved in the local frame of the scatterer.
Below we report the ``linear'' (i) and ``nonlinear'' (ii) implementation of the polarization source term.

{(i) - \emph{``Linear source term''.  }} For the numerical results shown in Fig.~\ref{fig:esempio} and in Fig.~\ref{fig:nonlin}(c)-(d), the polarization source term is implemented as a linear relativistic inhomogeneity (RI), travelling at velocity $v$ along the $z$ direction, i.e.,
\begin{equation}
P(z,t)=\epsilon_0 \{[n_0 + \delta n(t-z/v)]^2-n_0^2 \}E(z,t),
\end{equation}
where the background term, $n_{0}$, is the refractive index at the input carrier frequency and $\tau=t-z/v$ is the local time coordinate.
The ``artificial'' RI ($\delta n$) has a super-Gaussian longitudinal profile of amplitude $\delta n_{0}$ and thickness $\sigma$: $\delta n(\tau) = \delta n_{0}\exp[-(\tau-t_{0})^{m}/\sigma^{m}]$; being $m$ a positive even number identifying the Gaussian order and $t_{0}$ the initial temporal position.

{(ii) - \emph{``Nonlinear source term''. }} The numerical model is extended to include the nonlinear case presented in Fig.~\ref{fig:nonlin}(a)-(b); that is the case in which the inhomogeneity is induced by an intense soliton through the nonlinear Kerr effect. The polarization source term is $P(z,t)=\varepsilon_0  n_0n_2 |\mathcal{E}(z,t)|^2 E(z,t)$  
where $\mathcal{E}$ is the complex valued analytical signal and  $(1/2)n_2 |\mathcal{E}(z,t)|^2$ gives the dimensionless refractive index variation.\\ 
{ The probe pulse scattering by the soliton-induced RI, is modeled by solving a second equation for the real field of the probe pulse, $E_{2}(z,t)$, coupled to the equation for the soliton (\ref{eq:model}):
\begin{equation}
\partial_{z}{E_{2,\omega}}-i[k_{z}(\omega)-\omega/v]{E_{2,\omega}} = i\frac{\omega^{2}}{2 \epsilon_0 c^{2} k_{z}(\omega)}\, P_{2,\omega}.
\label{eq:model1}
\end{equation}
The coupling is indeed given by the ``cross-phase modulation'' term of the nonlinear polarization, i.e., $P_{2}(z,t)=2\varepsilon_0  n_0n_2 |\mathcal{E}(z,t)|^2 E_{2}(z,t)$, where $\mathcal{E}$ is the complex valued analytical signal of the pump (soliton), as in the previous case. 
}

We intentionally neglect all the other nonlinear terms, e.g. third harmonic generation and Raman scattering, in order to highlight the generation of the RR and NRR modes. 
Moreover, we recall that, in order to capture the coupling between positive and negative frequencies, the model equation (and especially the source term), must be implemented by calculating the $\omega$-frequency component of the \emph{real} polarization $P(z,t)$. This is achieved for each propagation step by means of back and forth Fourier transforms allowing for an evaluation of the real electric field $E(z,t)$ from its complex spectral components $E_\omega(z)$. 



\newpage
\clearpage

%
%
%
%

\section{Additional Material}

\subsection{A: Dispersion relations.}
As in all the scattering processes, the input and output scattered modes described in the main text, should obey precise momentum conservation laws. In the case of light scattered from a travelling inhomogeneity, the wave-vector $k(\omega_{\rm{RR}})$ and frequency $\omega_{\rm{RR}}$ of the scattered wave are linked to the input pulse by the condition:
\begin{equation}
k(\omega_{\rm{RR}})=k(\omega_{\rm{IN}})+\frac{\omega_{\rm{RR}}-\omega_{\rm{IN}}}{v},
\label{eq:momentum}\tag{S1}
\end{equation}
where $k(\omega)=\omega n(\omega)/c$ is the medium dispersion relation, $v$ the relativistic inhomogeneity (RI) velocity, and $\omega_{\rm{IN}}$ the input carrier frequency. The last term in Eq.~(\ref{eq:momentum}), $(\omega_{\rm{RR}}-\omega_{\rm{IN}})/v$, is related to the momentum transferred to the incoming pulse from the RI \cite{faccio}. This relation is rigorously derived in the following section.\\
In order to understand the origin of the negative RR, it is useful to find its coordinates on the dispersion curve $k(\omega)$ by graphically solving Eq.~(\ref{eq:momentum}). Without any loss of generality, in Figure~\ref{fig:dispersion} we consider a simple dispersion relation such as that of diamond, which does not exhibit any resonances over a very broad bandwidth (from less than one Terahertz through to ultraviolet wavelengths).  
\begin{figure}[b]
\centering
\includegraphics[scale=1]{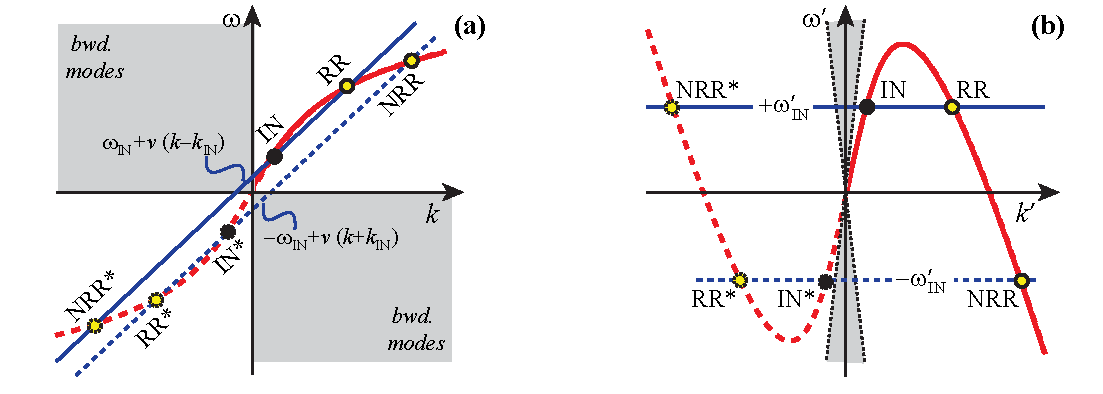}
\caption{(a) Dispersion relation of diamond in the lab. frame ($\omega,k$) and (b) in the reference frame ($\omega^{\prime},k^{\prime}$) moving at velocity $v$. Solid (dashed) red curves correspond to positive (negative) dispersion branch in the lab. frame. Straight blue lines indicate the momentum conservation condition, Eq.~(\ref{eq:momentum}) (solid blue), and its ``phase-conjugate'' counterpart (dashed blue). Grey shaded areas indicate the domains of backward propagating modes in the lab. frame, which are not excited by a forward propagating input pulse and are not considered in this work.}
\label{fig:dispersion}
\end{figure}
Equation~(\ref{eq:momentum}) is a straight line when represented in $(\omega,k)$ coordinates, as in Fig.~\ref{fig:dispersion}(a) (blue solid line). The frequency and wave-vector values of the propagating modes are thus determined by the overlap of this line with the actually allowed modes in the medium, i.e. by the intersection with the medium dispersion relation (red solid curve). Red dashed curves indicate the negative frequency branch (in the lab. frame) of the dispersion relation, while the grey shaded areas indicate the domains of the backward propagating modes (namely, the $2$nd and $4$th quadrants), which are not being excited by the forward propagating input mode and are not considered in this work. 
A first intersection is identified as the incoming mode IN. A second mode can be found, RR to which light may be resonantly scattered. Although the generation of resonant radiation is typically considered to be a property of self-scattering from intense nonlinear soliton pulses, we note that actually it is a more general process that only requires the existence of some kind of inhomogeneity in a \emph{linear} medium, as discussed here. A third intersection is found on the negative frequency branch of the dispersion curve: this is the negative frequency resonant radiation mode, NRR*.  Light may be resonantly transferred to this mode too, alongside the RR mode, as observed in recent experiments \cite{NRR}. 
We stress that the existence of a solution at negative frequencies is not unphysical, being the electromagnetic field a \emph{real}-valued quantity, that naturally oscillates at positive and negative frequencies. 
Indeed, the mode that is actually measured in an experiment is obtained by adding the three modes described above with their complex conjugate counterparts. These are found from the intersections of the dispersion curve with the straight dashed line in Fig.~\ref{fig:dispersion}(a), obtained from Eq.~\eqref{eq:momentum} by substituting $\omega$ with $-\omega$ and using the fact that the dispersion relation is an odd function, i.e. $k(-\omega)=-k(\omega)$.
Physically, this describes the momentum conservation condition in which the complex conjugate of the IN mode, IN*, is being scattered towards RR* and NRR.  \\
In Fig.~\ref{fig:dispersion}(b) we then consider the same dispersion relation in the reference frame that is comoving with the inhomogeneity, i.e. in $(\omega^\prime,k^\prime)$ given by the Doppler transformations: {$\omega^{\prime}=\gamma(\omega-vk)$} and {$k^{\prime}=\gamma(k-\omega v/c^{2})$}, where $\gamma=1/\sqrt{1-(v/c)^{2}}$ and $c$ is the speed of light in vacuum. In this reference frame the momentum conservation relation (S1) transforms into a condition on the comoving frequencies, $\omega^\prime=\omega{^\prime}_{\rm{IN}}$, and therefore represented as a horizontal line in the graph. Specifically, we find  $\omega_{\rm{RR}}^{\prime}=+\omega_{\rm{IN}}^{\prime}$ and $\omega_{\rm{NRR}^*}^{\prime}= + \omega_{\rm{IN}}^{\prime}$. The  NRR mode thus satisfies  $\omega_{\rm{NRR}}^{\prime}= -\omega_{\rm{IN}}^{\prime}$. 
 The RR (NRR) mode is thus naturally described by a  \emph{positive} (\emph{negative}) frequency in the comoving reference frame, reflecting the fact that RR (NRR) builds from frequencies with positive (negative) local phase velocities. 
We report here also the Sellmeier equation for the dispersion relation of diamond used in our simulations: $n^{2}(\lambda) =  1 + 4.3356\,\lambda^2/(\lambda^2-0.1060^2) + 0.3306\,\lambda^2/(\lambda^2-0.1750^2)$, where the wavelength $\lambda=2\pi c/\omega$ is expressed in $\mu$m.

\subsection{B: First Born approximation.}
In scattering theory the \emph{linear} relation involving input and output modes may be modeled with the scattering matrix $\mathbf{S}$, which is a unitary matrix connecting asymptotic states and which describes all the scattering channels of the process.
Physically the square moduli $|S_{j,i}|^{2}$ of the $\mathbf{S}$-matrix elements represent the number of photons for the $j-$output mode, which is being scattered by the $i-$input mode.
Following Kolesik \emph{et al.} \cite{kolesik-2010}, we evaluate the elements of the $\mathbf{S}$-matrix in the first Born approximation, giving an analytical formulation for the spectral amplitude of the scattered wave.
We also generalize the treatment in order to include the possibility of a third output scattered mode, i.e. the NRR mode that, as argued above may also be generated during the scattering process.\\
The role of the ``scattering potential'', $R(z,t)$, is played by the moving relativistic inhomogeneity (RI), which we assume to be stationary along propagation, i.e. $R(z,t)=R(t-z/v)=\delta n(t-z/v)$.\\
In detail, let us consider the laser pulse evolution ``unidirectional pulse propagation equation'' (UPPE) in the laboratory reference frame:
\begin{equation} 
\partial_z E_\omega -ik_z(\omega) E_\omega =i \frac {\omega^2}{2\epsilon_0 c^2 k_z(\omega)} P_\omega,
\label{uppe}\tag{S2}
\end{equation}
where
\begin{equation}\tag{S3}
P(z,t)=\epsilon_0 \left\{ [n_0+\delta n(t-z/v)]^2-n_0^2 \right\} E(z,t) \simeq 2\epsilon_0 n_0 \delta n (t-z/v) E(z,t),
\end{equation}
so that
\begin{equation}\tag{S4}
P_\omega =\frac 1{2\pi} \int dt\ 2\epsilon_0 n_0 \delta n (t-z/v) E(z,t) e^{i\omega t}.
\end{equation}
Equation~(\ref{uppe}) is the so-called ``unidirectional pulse propagation equation'' described in the Methods section and also numerically solved in this work.
We want to solve this equation perturbatively by writing
\begin{equation}\tag{S5}
E_\omega(z)=E_\omega^{(0)}(z)+E_\omega^{(1)}(z)+E_\omega^{(2)}(z)+\ldots,
\end{equation}
where $E_\omega^{(l)}(z)$ is of order $(\delta n)^l$. We then get the equations:
\begin{align}
& \partial_z E^{(0)}_\omega -ik_z(\omega) E^{(0)}_\omega=0, \label{zero} \tag{S6}\\
& \partial_z E^{(1)}_\omega -ik_z(\omega) E^{(1)}_\omega= i \frac {\omega^2 n_0}{c^2 k_z(\omega)}
\frac 1{2\pi} \int dt\ \left[\delta n (t-z/v) e^{i\omega t} \int e^{-i\omega' t} E_{\omega'}^{(0)}(z) d\omega'\right],\tag{S7} \label{uno} \\ \nonumber
& \vdots
\end{align}
From (\ref{zero}) we get
\begin{equation}\tag{S8}
E_\omega^{(0)}(z)=e^{ik_z(\omega) z} E_\omega^0,
\end{equation}
where $E_\omega^0$ is constant. Now, we can solve Eq.~(\ref{uno}) and, assuming that for $z\to-\infty$ there is only the unperturbed solution, we find,
\begin{equation}\tag{S9}
E_\omega^{(1)}(z)=i \frac {\omega^2 n_0}{2\pi c^2 k_z(\omega)} e^{ik_z(\omega) z} \int_{-\infty}^z d\xi e^{-ik_z(\omega) \xi}
\left\{
\int \delta n (t-\xi/v) e^{i\omega t} \int e^{-i\omega' t+ik_z(\omega')\xi}\, E_{\omega'}^0\, dt\, d\omega'
\right\}.
\end{equation}
We can rewrite this as
\begin{equation}\tag{S10}
E_\omega^{(1)}(z)=\int d\omega' \sigma(z,\omega,\omega') E_{\omega'}^{(0)}(z) \label{conv},
\end{equation}
where $\sigma$ is the $\mathbf{S}$-matrix density:
\begin{equation}\tag{S11}
\sigma(z,\omega,\omega_{\textrm{IN}})=i \frac {\omega^2 n_0}{2\pi c^2 k_z(\omega)} e^{i[k_z(\omega)-k_z(\omega_{\textrm{IN}})]z}
\int dt \int_{-\infty}^z d\xi e^{i[\omega t-k_z(\omega) \xi]} \delta n (t-\xi/v) e^{-i[\omega_{\textrm{IN}} t-k_z(\omega_{\textrm{IN}})\xi]}.
\end{equation}
By using the new set of variables, $u=t-\xi/v$ and $w=t+\xi/v$, we can evaluate the integrals in the last expression, to obtain:
\begin{align}
\sigma(z,\omega,\omega_{\textrm{IN}}) & \approx i \frac {v\omega^2 n_0}{2c^2 k_z(\omega)} e^{i[k_z(\omega)-k_z(\omega_{\textrm{IN}})]z} \times \nonumber \\
 & \times \delta \left\{ \frac{}{} \omega -\omega_{\textrm{IN}} - v \cdot [k_z(\omega)-k_z(\omega_{\textrm{IN}})]\right\} \times 
\hat R \left\{\frac 12 \left[\frac{}{} \omega -\omega_{IN} +v\cdot [k_z(\omega)-k_z(\omega_{\textrm{IN}})] \right] \right\},\label{sigma}\tag{S12}
\end{align}
where $\delta$ is the Dirac delta function, which is nonzero only for $\omega-\omega_{\rm{IN}}-v[k_{z}(\omega)-k_{z}(\omega_{\rm{IN}})]=0$, and $\hat{R}$ is the Fourier transform of the scattering potential $\delta n(u)$,
\begin{equation}\tag{S13}
\hat R(\omega):=\int du\, \delta n(u) e^{i\omega\, u}.
\end{equation}
We can now substitute (\ref{sigma}) in Eq.~(\ref{conv}) and by performing the integration (using the delta function) we finally obtain:
\begin{equation}\tag{S14}
E_\omega^{(1)}(z) =S(z,\omega, \omega_{\textrm{IN}}) E_{\omega_{\textrm{IN}}}^{(0)}(z),
\end{equation}
where the $\mathbf{S}$-matrix element is:
\begin{equation} \label{S}\tag{S15}
S(z,\omega, \omega_{\textrm{IN}}) = i \frac {v\omega^2 n_0}{2c^2 k_z(\omega)} e^{i(\omega-\omega_{\textrm{IN}})z/v} \hat R (\omega-\omega_{\textrm{IN}}),
\end{equation}
and
\begin{equation}\tag{S16}
\frac {\omega-\omega_{\textrm{IN}}}v=k_z(\omega)-k_z(\omega_{\textrm{IN}}).
\end{equation}
By substituting the delta function condition for $\omega$ spanning over positive and negative values, we then retrieve Equations~(1) and (2) of the main text.\\
Summarizing, the scattering matrix elements that describe the scattering of a laser pulse from a RI, may be explicitly evaluated from Eq.~(\ref{S}): the amplitude of the scattered wave is directly proportional to the Fourier transform of the scattering potential, i.e. of the RI. Moreover, Eq.~(\ref{eq:momentum}) emerges as natural condition that has to be satisfied in order for the scattering process to occur. This therefore provides a firm grounding for the fundamental momentum conservation relation used in this work, Eq.~(\ref{eq:momentum}).

\subsection{C: Generalized Manley-Rowe relation.}
For the scattering process under study, starting from an input seed pulse at frequency $\omega_{\rm{IN}}$, the photon number of the output waves is linked to the photon number of the input mode, $|\rm{IN}|^{2}$, by the following Manley-Rowe relation, generalized to the case of a moving scatterer \cite{ostrovskii,kravtsov-ostrovskii-stepanov}:
\begin{equation}
\textrm{sign}{(\gamma_{\rm{RR}})} |\rm{RR}|^{2}+\textrm{sign}{(\gamma_{\rm{NRR}})} |\rm{NRR}|^{2}=\textrm{sign}{(\gamma_{\rm{IN}})},\tag{S17}
\label{eq:gen-mr1}
\end{equation}
where $|{\rm{RR}}|^{2}$ and $|{\rm{NRR}}|^{2}$ are the photon numbers of the RR and NRR modes normalized to the input photon number, $|{\rm{IN}}|^{2}$, and $\gamma_i = (v_{\phi}^2- v v_{\phi})_i$,  where $v_{\phi}$ is the phase velocity of the $i$-mode. We note that $\textrm{sign}{(\gamma_i)}=\textrm{sign}{(\omega_i-v k_i)}=\textrm{sign}{(\omega_i')}$; while the real-valued RR mode will have positive frequency in the comoving frame, i.e. $\omega_{\rm{RR}}'>0$, the real-valued NRR mode will always have a negative comoving frequency, $\omega_{\rm{NRR}}'<0$.
Therefore, Eq.~(\ref{eq:gen-mr1}) assumes the form of Eq.~(3), that is $|\rm{RR}|^{2}-|\rm{NRR}|^{2}=1$.

\subsection{D: Propagation Cross-Correlation.}
We evaluate the propagation cross-correlation map for the numerics presented in the main text, by adopting the approach of B\'ejot \emph{et al.} in \cite{bejot}. This correlation is positive if frequencies are created simultaneously, i.e. at the same propagation distance $z$, and negative if one frequency is used to create the other. For each pair of frequencies the propagation cross-correlation is defined as:
\[
C(\omega_{1},\omega_{2})= \frac{\textrm{Cov}\left\{ I(\omega_{1}),I(\omega_{2})\right\}}{\sqrt{ \textrm{Cov}\left\{I(\omega_{1}),I(\omega_{1})\right\}\textrm{Cov}\left\{I(\omega_{2}),I(\omega_{2})\right\}}},\tag{S18}
\]
where, 
\[
\textrm{Cov}\left\{ I(\omega_{1}),I(\omega_{2})\right\}= \langle [I(\omega_{1},z)-\langle I(\omega_{1},z)\rangle]\,[I(\omega_{2},z)-\langle I(\omega_{2},z)\rangle]\rangle,\tag{S19}
\]
where $\langle \cdot \rangle$ indicates the expectation value i.e. the average evaluated over the propagation distance $z$ 
Figure~\ref{fig:corr}(a) shows the propagation cross-correlation map for the numerical results reported in Fig.~2 of the main paper, which refer to a \emph{linear} scattering in diamond. The two resonant modes, RR and NRR, are perfectly correlated (correlation equal to $+1$), implying that they are born together, whereas they are both anti-correlated with the input mode. This confirms that indeed the RR and NRR modes are the result of the same scattering event seeded by the IN mode. Finally, Figure \ref{fig:corr}(b) shows the propagation cross-correlation map for the numerics of Fig.~3(a)-(b) of the main paper, referring to a \emph{nonlinear} scattering in fused silica. Here too, the map shows obvious signs of correlation between the resonant modes.
\begin{figure}[t]
\centering
\includegraphics[scale=1]{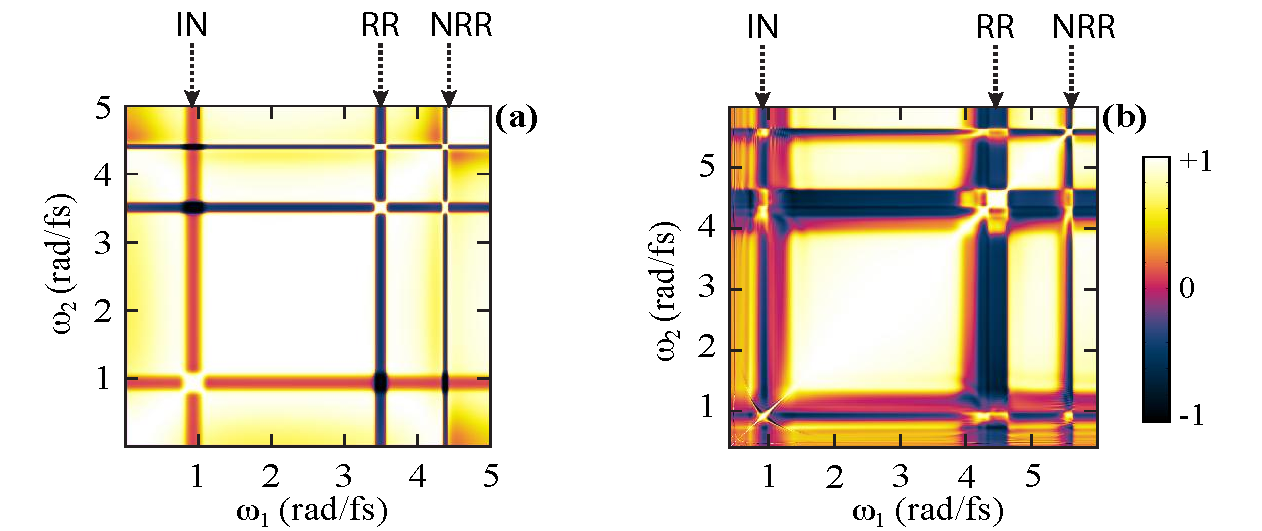}
\caption{Propagation cross-correlation maps for (a) \emph{linear} scattering in diamond and (b) \emph{nonlinear} scattering in fused silica. The two maps refer to the numerics presented in the main paper, Figures 2 and 3(a)-(b), respectively. Arrows indicate the position of the IN, RR and NRR modes.}
\label{fig:corr}
\end{figure}

\subsection{E: Comparison between a soliton-induced and a linear, ``toy-model'', RI.}
Figure \ref{fig:nonlin} is an extension of Figure 3 of the main paper and shows a comparison between the dynamics of a soliton-induced RI, with self-scattering towards the resonant modes (a)-(b), and an artificial, ``toy-model'' RI co-propagating with a weak probe pulse in a purely linear medium (c)-(d). Details of the simulations are indicated in the main text.
\begin{figure}[t]
\centering
\includegraphics[scale=1]{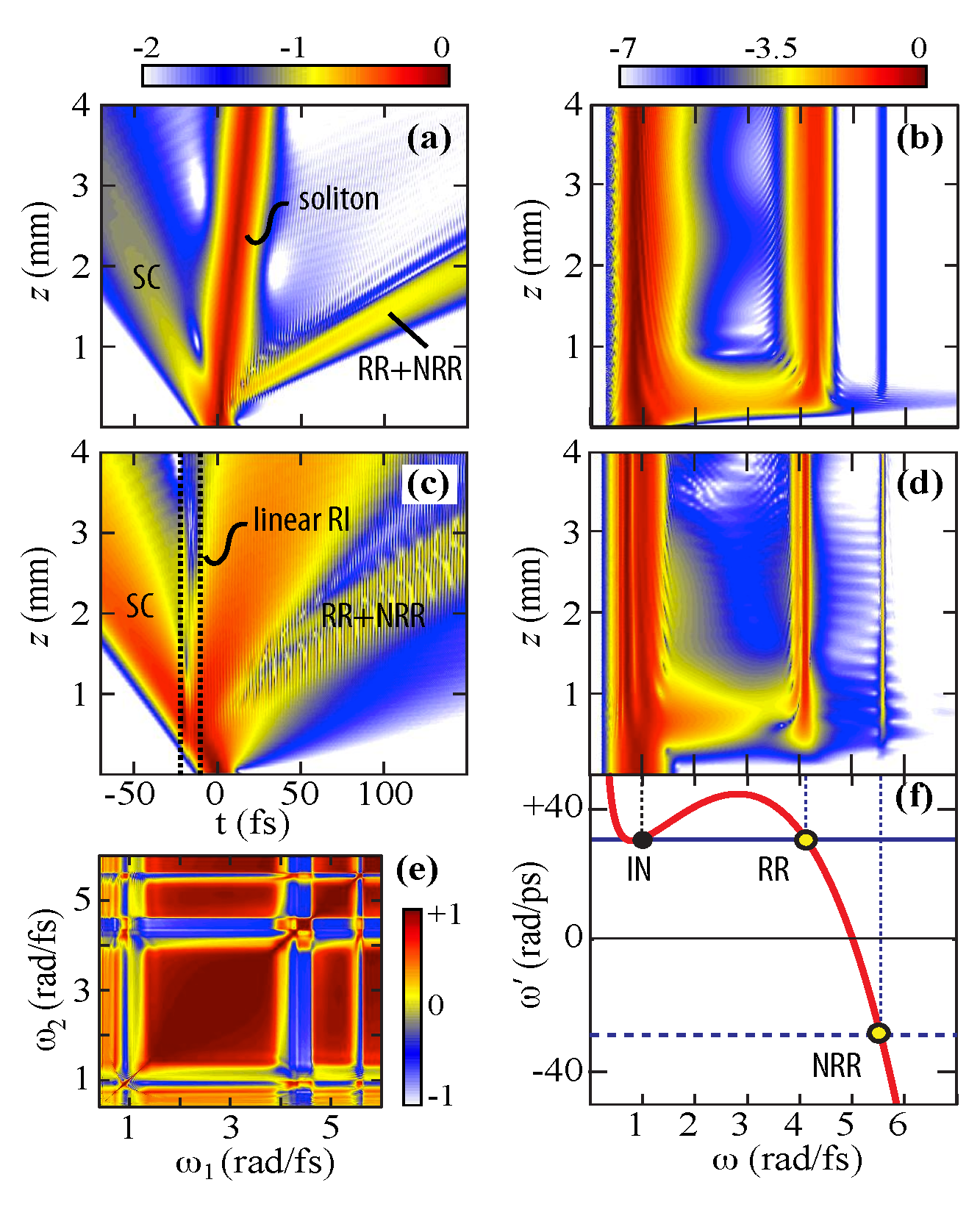}
\caption{
RI scattering dynamics in fused silica. Comparison between a soliton-induced (via the \emph{nonlinear} Kerr effect) RI (a)-(b) and the ``toy-model'' RI with the same shape as the soliton-induced inhomogeneity, in a purely linear medium (c)-(d). Envelope (left hand side), and spectral (right hand side) evolution in propagation. 
Propagation cross-correlation map for the nonlinear case (e). 
Theoretical positions of input (IN) and resonant (RR, NRR) modes on the comoving dispersion curve $\omega^{\prime}=\omega^{\prime}(\omega)$, indicated with filled circles (f). }
\label{fig:nonlin}
\end{figure}

For both cases the frequencies of the RR and NRR modes are in perfect agreement with the theoretical predictions, as shown on the comoving dispersion curve ($\omega,\omega^{\prime}$) in Fig.~\ref{fig:nonlin}(f).
Finally, Figure \ref{fig:nonlin}(e) [as Fig.~\ref{fig:corr}(b)] shows the propagation cross-correlation for the nonlinear case, as described in detail in Section D and in \cite{bejot}. The two resonant modes are correlated (correlation  equal to  $+1$) implying that they are born together, whereas they are both anti-correlated (correlation  equal or close to $-1$) with the input mode. This provides further confirmation that indeed the RR and NRR modes are the result of the same scattering process seeded by the IN mode.

\newpage



\end{document}